\RequirePackage{fix-cm}
\documentclass[english,italian,american,british,11pt]{article}
\usepackage[T1]{fontenc}
\usepackage{xcolor}
\usepackage{pdfcolmk}
\usepackage{textcomp}
\usepackage{amsthm}
\usepackage{amsmath}
\usepackage{amssymb}
\usepackage{setspace}
\usepackage{dsfont}
\usepackage{esint}
\usepackage{floatrow}
\usepackage{babel}
\usepackage[latin9]{inputenc}
\usepackage{graphicx}
\usepackage{amsthm}
\usepackage{amssymb}
\usepackage{hyperref}
\usepackage{fancyhdr}
\usepackage{nonfloat}
\PassOptionsToPackage{normalem}{ulem}
\usepackage{ulem}
\makeatletter

\numberwithin{equation}{section}
\numberwithin{figure}{section}
\numberwithin{table}{section}
  \theoremstyle{plain}
  
  \theoremstyle{definition}
  
  \theoremstyle{plain}
  \newtheorem{thm}{\protect\theoremname}[section]
  \theoremstyle{plain}
  \newtheorem{lemma}{\protect\lemmaname}[section]
  
\@ifundefined{date}{}{\date{}}
\usepackage[all]{xy}
\usepackage{xcolor}

\input xy
\xyoption{all}
\xyoption{frame}
\xyoption{tile}

\makeatother

\usepackage{babel}
  \addto\captionsamerican{\renewcommand{\definitionname}{Definition}}
  \addto\captionsamerican{\renewcommand{\propositionname}{Proposition}}
  \addto\captionsenglish{\renewcommand{\definitionname}{Definition}}
  \addto\captionsenglish{\renewcommand{\propositionname}{Proposition}}
  \addto\captionsitalian{\renewcommand{\definitionname}{Definizione}}
  \addto\captionsitalian{\renewcommand{\propositionname}{Proposizione}}
  \providecommand{\definitionname}{Definition}
  \providecommand{\propositionname}{Proposition}
  \providecommand{\lemmaname}{Lemma}
\addto\captionsamerican{\renewcommand{\theoremname}{Theorem}}
\addto\captionsenglish{\renewcommand{\theoremname}{Theorem}}
\addto\captionsitalian{\renewcommand{\theoremname}{Teorema}}
\providecommand{\theoremname}{Theorem}

\let\l=\lambda
\let\p=\pi
\let\d=\delta
\let\e=\varepsilon
\let\m=\mu
\let\O=\Omega
\let\f=\varphi
\let\c=\chi
\let\x=\xi
\let\dpr=\partial
\let\L=\Lambda
\let\G=\Gamma
\let\t=\tau
\let\a=\alpha

\def\xx{{\bf x}}
\def\qq{{\bf q}}

\def\be{\begin{equation}}
\def\ee{\end{equation}}
\def\bea{\begin{eqnarray}}
\def\eea{\end{eqnarray}}

\begin{document}

\title{\textbf{Construction of the Lyapunov spectrum in a chaotic system displaying phase
synchronization}}

\author
{Leonardo De Carlo$^1$, Guido Gentile$^2$, Alessandro Giuliani$^2$
\vspace{2mm}
\\ \small
$^{1}$ Gran Sasso Science Institute (GSSI), Viale Francesco Crispi 7, L'Aquila, 67100, Italy
\\ \small 
$^{2}$ Dipartimento di Matematica e Fisica, Universit\`a di Roma Tre, Roma, 00146, Italy
\\ \small 
E-mail:  neoleodeo@gmail.com, gentile@mat.uniroma3.it, giuliani@mat.uniroma3.it}

\date{} 
 
\maketitle 

\begin{abstract}
We consider a three-dimensional chaotic system consisting of the suspension of Arnold's cat map
coupled with a clock  via a weak dissipative interaction. We show that the coupled system displays a synchronization
phenomenon,  in the sense that the relative phase between the suspension flow and the clock locks to a special value, 
thus making the motion fall onto a lower dimensional attractor.  More specifically, we construct the 
attractive invariant manifold, of dimension smaller than three, using a convergent perturbative expansion. 
Moreover, we compute via convergent series the Lyapunov exponents, including notably the central one.
The result generalizes a previous construction of the attractive invariant manifold in a similar but simpler model.
The main novelty of the current construction relies in the computation of the Lyapunov spectrum, which consists
of non-trivial analytic exponents. Some conjectures about a possible smoothening transition of the attractor as the coupling is increased are also discussed.

\vspace{2mm}

\end{abstract}

\section{Introduction} \label{sec:1}

Synchronization in chaotic systems is a surprising phenomenon, which recently received a lot of attention,
see e.g. \cite{ADKMZ, Blech,BKOVZ,Gonzalez,PCJMH,PRK}.
Even though the heuristic theory and the classification of the synchronization phenomena are well studied
and reasonably well understood, a mathematically rigorous theory is still lacking. Generally speaking, 
a standard difficulty lies in the fact that the phenomenon involves the dynamics of non-uniformly 
chaotic systems, typically consisting of different sub-systems, whose long-time behavior depends crucially
on the sign of the ``central'' Lyapunov exponents, i.e. of those exponents that are zero in the case of zero coupling, 
and become possibly non-trivial in the presence of interactions among the sub-systems. 
The mathematical control of such exponents is typically very hard. Progress in their computation is
a fundamental preliminary step for the construction of the SRB measure of chains or lattices of chaotic flows, 
which may serve as toy models for extensive chaotic systems out-of-equilibrium (i.e. they may serve as 
standard models for non-equilibrium steady states in non-equilibrium statistical mechanics).

In a previous paper \cite{GGG}, we introduced a simple model for phase synchronization in a three-dimensional system
consisting of the suspension flow of Arnold's cat map coupled with a clock. The coupling in \cite{GGG} was unidirectional,
in the sense that it did not modify the suspension flow, but only the clock motion. Notwithstanding its simplicity, the model has 
a non-trivial behavior: in particular, it exhibits phase locking and in \cite{GGG} we constructed the corresponding attractive invariant manifold via a 
convergent expansion. However, because of unidirectionality, the Lyapunov spectrum in \cite{GGG} was very simple:
the ``longitudinal'' exponents (i.e., those corresponding to the motion on the invariant manifold)
coincided with the unperturbed ones, and
the central exponent was expressed in the form of a simple integral of the perturbation over the manifold.
In this paper, we extend the analysis of \cite{GGG}  to a simple bidirectional model, for which the Lyapunov spectrum is 
non-trivial, and we show how to compute it in terms of a modified expansion, which takes the form of a
decorated tree expansion discussed in detail in the following. 

The model is defined as follows. Take Arnold's cat map $x \mapsto S x$ and denote by $\lambda_{\pm}$ and $x_{\pm}$
the eigenvalues and eigenvectors, respectively, of $S  $:
\begin{equation} \label{eq:1.1}
S  = \left(\begin{array}{cc} 2 & 1\\ 1 & 1 \end{array}\right) , \qquad
\lambda_{\pm}=\frac{3\pm\sqrt{5}}{2} , \qquad
x_{\pm}=\pm\frac1{\mathcal N_\pm}\begin{pmatrix} 1\\ \l_\pm-2\end{pmatrix}\,,
\end{equation}
with $\mathcal N_\pm:=\sqrt{1+(\l_\pm-2)^2}$, so that $x_\pm$ are normalized. We let the suspension flow of Arnold's cat 
be defined as $x \to x(t):=\Phi^t(x)$, with $x(t)=S^{k}x$, if $2\p k< t\le 2\p (k+1)$ $\forall k\in\mathbb Z$. Formally, 
$x(t)$ is the solution to the following differential equation\footnote{In \cite{GGG} we erroneously wrote $\dot x=\d(t) (S  -1)x$ instead of \eqref{1.susf},
but throughout the paper we only used the fact that at all times $2\p k+0^+$ the variable $x$ jumped
abruptly from $x(2\p k)$ to $S   x(2\p k)$, and besides these discontinuities the flow was smooth.
Therefore, all the results and statements of \cite{GGG} are correct, modulo this re-interpretation of
the flow equation \cite[(2.1)]{GGG}, where $\d(t)(Sx-x)$ should be replaced by $\d(t)(\log S)x$.} on $\mathbb T^2=\mathbb R^2/(2\p \mathbb Z)$:
\be \dot x=\d(t)\big(\log S\big) x\,, \label{1.susf}\ee
where $\delta$ is the $2\pi$-periodic delta function such that $\int_{0}^{t}dt'\delta(t')=1$ for all
$0<t\leq2\pi$.
The model of interest is obtained by coupling the suspension flow of Arnold's cat map with a clock
by a regular perturbation, so that on $\mathbb{T}^{2}\times\mathbb{T}$ the evolution equation is
\begin{equation} \label{eq:1.2}
\begin{cases} 
\dot{x}=\delta(t)\big[\log S\big]x+\varepsilon f(x,w,t) , & \\
\dot{w}=1+\varepsilon g(x,w,t),
\end{cases}
\end{equation}
where $\varepsilon \ge 0$ and $f$, $g$ are $2\pi$-periodic in their arguments. For $\e=0$ the 
motions of $x$ and $w$ are independent. Therefore, the relative phase $w$ mod $2\p$ among the two flows 
is arbitrary. If $\e\neq0$ and if the interaction is dissipative (in a suitable sense, to be clarified in a moment),
then the phases of the two sub-systems can lock, so that the limiting motion in the far future takes place on 
an attractor of dimension smaller than 3, for all initial data in an open neighborood of the attractor.
In  \cite{GGG}, we explicitly constructed such an attractor in terms of a convergent power series expansion in $\e$,
for $f=0$ and a special class of dissipative functions $g$. In this paper, we generalize the analysis of \cite{GGG} 
to $f\neq 0$. Our first result concerns the construction of the attractive invariant manifold for $f\neq 0$.

\begin{thm} \label{prop:1}
Let $S_{\varepsilon}^{t}$ be the flow on $\mathbb{T}^{3}$
associated with the dynamics \eqref{eq:1.2}, with $f$ and $g$ analytic in their arguments. Set
\begin{equation} \label{eq:1.3}
\bar{\gamma}_{0}(\varphi,w_{0}):= \int_{0}^{2\pi} {\rm d}t \, g(S \varphi,w_{0}+t,t), \qquad
\bar{\gamma}_{1}(\varphi,w_{0}):=\int_{0}^{2\pi}{\rm d}t \, \partial_{w}g(S \varphi,w_{0}+t,t),
\end{equation}
and assume there exists $w_{0}\in\mathbb{T}$ such that $\bar{\gamma}_{0}(\varphi,w_{0})=0$
and $\Gamma:=\bar{\gamma}_{1}(\varphi,w_{0})<0$, independently of $\varphi$.
Then there are constants $\varepsilon_{0},\, c>0$ such that for $0<\varepsilon<\varepsilon_{0}$
there exist a homemorphism $H:\mathbb{T}^{2}\rightarrow\mathbb{T}^{2}$ and a continuous function
$W:\mathbb{T}^{2}\rightarrow\mathbb{T}$,  both H\"older-continuous of exponent $\beta\geq c\varepsilon$,
such that the surface $\Omega=\left\{ (H(\varphi),W(\varphi)) : \varphi\in\mathbb{T}^{2}\right\}$
is invariant under the Poincar\'e map $S_{\varepsilon}^{2\pi}$ and the dynamics of $S_{\varepsilon}^{2\pi}$
on $\Omega$ is conjugated to that of $S $ on $\mathbb{T}^{2}$, i.e.
\begin{equation} \label{eq:1.4}
S_{\varepsilon}^{2\pi}(H(\varphi),W(\varphi))=(H(S \varphi),W(S \varphi)) .
\end{equation}
\end{thm}

The proof of this theorem is constructive: it provides an explicit algorithm for computing the generic term
of the perturbation series of $(H,W)$ with respect to $\e$, it shows how to estimate it and how to prove
convergence of the series. As a by-product, we show that the invariant manifold is holomorphic in $\e$ in a suitable domain 
of the complex plane, whose boundary contains the origin. The construction also implies that $\O$ is an attractor. 
We denote by $B$ its basin of attraction and by $B_0$ an arbitrary open neighborood of $\O$ contained 
in $B$ such that $\m_0(B_0)>0$, with $\m_0$ the Lesbegue measure on $\mathbb T^3$. 

In addition to the construction of the invariant surface, in this paper we show how to compute the invariant measure
on the attractor  and the Lyapunov spectrum, in terms of convergent expansions. More precisely, let $\m$
be the Lesbegue measure restricted to $B_0$, i.e., denoting by $\c_{B_0}$ the characteristic function
of $B_0$, $\m(A):=\m_0(\c_{B_0}A)$, for all measurable $A$. The ``natural'' invariant measure on the attractor,
$\m_{\rm SRB}$, is defined by
$$\lim_{t\to\infty}\frac1{t}\int_0^t F(S_\e^{\t}(\xx))d\t=\m_{\rm SRB}(F)$$
for all continuous functions $F$ and $\m$-a.e. $\xx$, where $\xx=(x,w)\in\mathbb T^3$. The limiting measure $\m_{\rm SRB}$ 
is supported on $\O$ and such that $\m_{\rm SRB}(\O):=\m_{\rm SRB}(\c_\O)=1$. On the attractor, $\m_{\rm SRB}$-a.e. point 
defines a dynamical base, i.e. a decomposition of the tangent plane as $T\mathbb T^3={\mathcal W}_+(\xx)
\oplus {\mathcal W}_0(\xx) \oplus {\mathcal W}_-(\xx)$, such that 
$$ \lim_{k\to\infty}\frac1{k}\log\frac{\big|\dpr S_\e^{2\p k}(\xx)d\x\big|}{|d\x|}=
\L_i\;,\qquad {\rm if}\qquad 0\neq d\x\in {\mathcal W}_i(\xx)\;,\quad i\in\{+,0,-\}\;. $$
The constants of motion $\L_i$ are the Lyapunov exponents, and we suppose them ordered as $\L_+>\L_0>\L_-$;
in the following we shall call $\L_0$ the \emph{central} Lyapunov exponent. Our second main result is the following.

\begin{thm} \label{prop:2} There exists $\e_0$ such that the following is true. 
Let $F$ be an H\"older continuous function on $\mathbb T^3$. Then $\m_{\rm SRB}(F)$ is H\"older continuous in $\e$, 
for $0<\e\le \e_0$. If $F$ is analytic, then $\m_{\rm SRB}(F)$ is analytic in $\e$, for $0<\e<\e_F\le \e_0$ and a suitable 
$F$-dependent constant $\e_F$. Moreover, the Lyapunov exponents $\L_i$, $i\in\{+,0,-\}$, are analytic in $\e$ 
for $0<\e\le \e_0$. In particular, the central Lyapunov exponent is negative: $\L_0=\e\G+O(\e^2)$,
while $\L_\pm=\log\l_\pm+O(\e)$.
\end{thm}

The paper is organized as follows. Theorem \ref{prop:1} is proved in Section \ref{sec:2} below.
The proof follows the same strategy of \cite{GGG}:  (1) we first write the equations for the invariant surface
and solve them recursively at all orders in $\e$; (2) then we express the result of the recursion 
(which is not simply a power series in $\e$) in terms of tree diagrams
(planar graphs without loops); trees with $n$ nodes are proportional to $\e^n$ times a {\it tree value}, 
which is also a function of $\e$; (3) finally, using the tree representation, we derive an upper bound on the tree values.
The fact that the dissipation is small, of order $O(\e)$, produces bad factors  $\e^{-k}$ in the bounds of the tree values,
for some $k$ depending on the tree. Therefore, we need to show that for any tree $k$ is smaller than a fraction of $n$,
if $n$ is the number of nodes in the tree. This is proved by exhibiting suitable cancellations, 
arising from the condition $\bar{\gamma}_{0}(\varphi,w_{0})=0$.

Theorem \ref{prop:2} is proved in Section \ref{sec:3}. The proof adapts the tree expansion to the 
computation of the local Lyapunov exponents $\L_i(\xx)$ on the invariant surface, in the spirit of \cite[Chapter 10]{GBG}
and \cite{BFG04}. The positive local Lyapunov exponent $\L_+(\xx)$ plays the role of the Gibbs potential
for the invariant measure $\m_{SRB}$. Therefore, given a convergent expansion for $\L_+(\xx)$, $\m_{SRB}$ 
can be constructed by standard cluster expansion methods, as in \cite[Chapter 10]{GBG}. Finally, $\L_i$ can be 
expressed as averages of the local exponents over the stationary distribution.

In Section \ref{sec:4}, we present some numerical evidences for a fractal to non-fractal transition
of the invariant manifold, and formulate some conjectures. 

\section{Formation of an invariant surface} \label{sec:2} 

\subsection{Conjugation}

In this section, we define the equations for the invariant manifold, by introducing a conjugation
that maps the dynamics restricted to the attractor onto the unperturbed one. The conjugation is denoted by
$(H,W):\mathbb{T}^{2}\rightarrow\left(\mathbb{T}^{2},\mathbb{T}\right)$, with
$$ H(\varphi):=\varphi+h(\varphi)=(\mathds{1}+h)(\varphi) , \qquad W(\varphi):=w_{0}+U(\varphi) , $$
where  $\mathds{1}$ is the identity in $\mathbb{R}^2$.

Let $x(0)=x=H(\varphi)$ and $w(0)=w_{0}+U(\varphi)$ be the initial conditions at time $t=0$.
We will look for a solution to \eqref{eq:1.2} of the form
\begin{equation} \label{eq:2.1}
\begin{cases}
x(t)=S x+a(\varphi,t)=S H(\varphi)+a(\varphi,t)=S \varphi+S h(\varphi)+a(\varphi,t) , & \\
w(t)=w_{0}+t+u(\varphi,t) , &
\end{cases}
\end{equation}
for $0< t\le2\pi$, with boundary conditions
\begin{equation} \label{eq:2.2}
a(\f,0^+)=0, \quad a(\f,2\p)=H(S \varphi)-SH(\f), \quad u(\f,0^+)=U(\f),\quad u(\varphi,2\pi)=U(S \varphi) .
\end{equation}
The evolution equation for $w$ will be written by ``expanding the vector field at first order in $u$ and at zeroth order in $S h+a$'', i.e. as
\begin{equation} \label{2.2bis}
\dot{u}(\varphi,t)=\varepsilon g(x(t),w(t),t) =\varepsilon\gamma_{0}(\varphi,t)+
\varepsilon\gamma_{1}(\varphi,t) \, u(\varphi,t)+\varepsilon G(\varphi,t),
\end{equation}
where 
\begin{equation} \label{eq:2.3}
\gamma_{0}(\varphi,t):=g(S \varphi,w_{0}+t,t) , \qquad
\gamma_{1}(\varphi,t):=\partial_{w}g(S \varphi,w_{0}+t,t) 
\end{equation}
and
\be G(\varphi,t):=g(S H(\varphi)+a(\varphi,t),w_0+t+u(\f,t),t)-\gamma_{0}(\varphi,t)-
\gamma_{1}(\varphi,t)\,u(\varphi,t).\label{2.G} \ee
The logic in the rewriting \eqref{2.2bis} is that the (linear) approximate dynamics obtained by neglecting $G$ is dissipative, 
with contraction rate proportional to $\e$, thanks to the second condition in \eqref{eq:1.3}: this will allow us to control
the full dynamics as a perturbation of the approximate one. The approximation obtained by neglecting $G$ is the simplest one
displaying dissipation. In principle we could have expanded the dynamics at first order both in $u$ and in $Sh+a$, 
but the result would be qualitatively the same. We now set
\begin{equation} \label{eq:2.4}
\Gamma(\varphi,t,\tau):=\int_{\tau}^{t} {\rm d}\tau' \, \gamma_{1}(\varphi,\tau') 
\end{equation}
and fix $w_0$ such that $\Gamma(\varphi,2\pi,0)=\Gamma<0$; then we obtain
\begin{equation} \label{eq:2.5}
u(\varphi,t)={\rm e}^{\varepsilon\Gamma(\varphi,t,0)}U(\varphi)+ \xi(\f,t) , \qquad
\xi(\f,t) := \varepsilon\int_{0}^{t}{\rm d}\tau \,
{\rm e}^{\varepsilon\Gamma(\varphi,t,\tau)} \left( \gamma_{0}(\varphi,\tau)+G(\varphi,\tau) \right) . 
\end{equation}
The equation for $x$, if expressed in terms of $a$, gives, after integration,
\begin{equation} \label{eq:2.6}
a(\f,t)=\varepsilon\int_{0}^{t} {\rm d} \tau \, f(SH(\f)+a(\f,\t),w_0+\t+u(\f,\tau),\tau).
\end{equation}

For $t=2\pi$, these give
\begin{subequations} \label{eq:2.7}
\begin{align}
U(S \varphi) & =e^{\varepsilon\Gamma}U(\varphi)+\varepsilon\int_{0}^{2\pi} {\rm d}\tau
\, {\rm e}^{\varepsilon\Gamma(\varphi,2\pi,\tau)}
\left( \gamma_{0}(\varphi,\tau)+G(\varphi,\tau) \right) , 
\label{eq:2.7a} \\
H(S \varphi) & =S H(\varphi)+\varepsilon\int_{0}^{2\pi}{\rm d} \tau \, f(SH(\f)+a(\f,\t),w_0+\t+u(\f,\tau),\tau).
\label{eq:2.7b}
\end{align}
\end{subequations}

It is useful to introduce an auxiliary parameter $\mu$, to be eventually set equal to $\e$,
and rewrite \eqref{eq:2.5} and \eqref{eq:2.7a} as
\begin{subequations} \label{eq:2.8}
\begin{align}
u(\varphi,t) & = {\rm e}^{\mu\Gamma(\varphi,t,0)}U(\varphi)+\x(\f,t)\;,\qquad \x(\f,t)=\varepsilon\int_{0}^{t}d\tau \,
{\rm e}^{\mu\Gamma(\varphi,t,\tau)}
\left( \gamma_{0}(\varphi,\tau)+G(\varphi,\tau) \right), 
\label{eq:2.8a} \\
U(S \varphi) & = {\rm e}^{\mu\Gamma}U(\varphi)+\varepsilon \int_{0}^{2\pi} {\rm d}\tau \,
{\rm e}^{\mu\Gamma(\varphi,2\pi,\tau)}
\left( \gamma_{0}(\varphi,\tau)+G(\varphi,\tau) \right).
\label{eq:2.8b}
\end{align}
\end{subequations}
The idea is to first consider $\mu$ as a parameter independent of $\varepsilon$, then write the solution
in the form of a power series in $\e$, with coefficients depending on $\m$,  and finally show that the
($\mu$-dependent) radius of convergence of the series in $\e$ behaves like $\m^\alpha$, $\alpha<1$, 
at small $\m$: this implies that we will be able to take $\m=\e$ without spoiling the summability of the series. 

Summarizing, we will look for a solution of \eqref{eq:1.2} in the form
\begin{equation} \label{eq:2.9}
x(t)=S \varphi+S h(\varphi)+a(\varphi,t), \quad w(t)= w_{0} + t + 
{\rm e}^{\mu \Gamma(\varphi,t,0)}U(\varphi)+\xi(\varphi,t), \end{equation}
with
\begin{subequations} \label{eq:2.11}
\begin{align}
\xi(\varphi,t) & = \varepsilon\int_{0}^{t} {\rm d}\tau \, {\rm e}^{\mu\Gamma(\varphi,t,\tau)}
\left( \gamma_{0}(\varphi,\tau)+G(\varphi,\tau) \right),
\label{eq:2.11a} \\
U(S \varphi) & = {\rm e}^{\mu\Gamma}U(\varphi)+\varepsilon\int_{0}^{2\pi} {\rm d}\tau \, 
{\rm e}^{\mu\Gamma(\varphi,2\pi,\tau)}
\left( \gamma_{0}(\varphi,\tau)+G(\varphi,\tau) \right),
\label{eq:2.11b} \\
a(\f,t) & = \varepsilon \int_{0}^{t} {\rm d}\tau \, F(\f,\t),
\label{eq:2.11c} \\
h(S \varphi) & =S h(\varphi)+\varepsilon\int_{0}^{2\pi} {\rm d}\tau \,
F(\f,\t).
\label{eq:2.11d}
\end{align}
\end{subequations}
and: $F(\f,t):=f(x(t),w(t),t)$, with $x(t)$ and $w(t)$ as in \eqref{eq:2.9};
$G$ is defined in \eqref{2.G}; $\m$ must be eventually set equal to $\e$.

\subsection{Recursive equations} \label{sec:2.2}

The solution to \eqref{eq:2.9}-\eqref{eq:2.11}  is looked for in the form of a power series expansion in $\varepsilon$
(at fixed $\m$, in the sense explained after \eqref{eq:2.8}). Therefore, we write 
\begin{subequations} \label{eq:2.10}
\begin{align}
& U(\varphi)=\sum_{n=1}^{\infty} \varepsilon^{n} U^{(n)}(\varphi) ,
\qquad \xi(\varphi,t) = \sum_{n=1}^{\infty} \varepsilon^{n}\xi^{(n)}(\varphi,t),
\label{eq:2.10a} \\
& h(\varphi)= \sum_{n=1}^{\infty} \varepsilon^{n}h^{(n)}(\varphi) ,
\qquad a(\varphi,t) = \sum_{n=1}^{\infty} \varepsilon^{n} a^{(n)}(\varphi,t),
\label{eq:2.10b}
\end{align}
\end{subequations}
and insert these expansions into \eqref{eq:2.11}.
By the analyticity assumption on $f$ and $g$, we may expand (defining $\qq=(q_+,q_-,q_3)$
and $\mathbb N_0= \mathbb Z\cap[0,+\infty)$)
\begin{eqnarray} 
& & \hskip-.6cm G(\varphi,t) = \sideset{}{^*}\sum_{\qq\in\mathbb N_0^3} G_{\qq}(\varphi,t)
\bigl(\l_+h_+(\f) +a_+(\f,t)\bigr)^{q_+} \bigl(\l_-h_-(\f)+a_-(\f,t) \bigr)^{q_-}
(e^{\mu\Gamma(\varphi,t,0)}U(\varphi)+\xi(\varphi,t))^{q_3}, \nonumber \\
& & \hskip-.6cm F(\varphi,t)=\sum_{\qq\in\mathbb N_0^3}
F_{\qq}(\varphi,t) 
\bigl(\l_+h_+(\f) +a_+(\f,t) \bigr)^{q_+} \bigl(\l_-h_-(\f)+a_-(\f,t) \bigr)^{q_-}
\bigl( {\rm e}^{\mu\Gamma(\varphi,t,0)}U(\varphi)+\xi(\varphi,t) \bigr)^{q_3}, \nonumber
\end{eqnarray}
where $*$ in the first sum denotes the constraint $\qq \neq (0,0,0),(0,0,1)$, and
$h_\pm=h\cdot x_\pm$ and similarly for $a_\pm$ (recall that $\l_\pm$ and $x_\pm$ are
the eigenvalues and eigenvectors of $S$); here and henceforth we are denoting by $\cdot$
the standard scalar product in $\mathbb R^2$. Moreover
\begin{equation} \nonumber
G_{\qq}(\varphi,t) :=\frac{1}{\qq!}\partial_{+}^{q_+}\dpr_-^{q_-}\partial_{w}^{q_3}g(S \varphi,w_{0}+t,t) ,
\qquad
F_{\qq}(\varphi,t) :=\frac{1}{\qq!}\partial_{+}^{q_+}\dpr_-^{q_-}\partial_{w}^{q_3}f(S \varphi,w_{0}+t,t) ,
\end{equation}
where $\qq!:=q_+!q_-!q_3!$ and $\dpr_\pm:=x_\pm\cdot\dpr_x$. For future reference,
we note since now that the analyticity of $f$ and $g$ yields, by the Cauchy inequality,
\begin{equation} \label{2.an}
|G_{\qq}|, |F_{\qq}|
\leq C_{0}^{|\qq|}  ,
\end{equation}
for some constant $C_{0}>0$, uniformly in $\f,t$ (here $|\qq|=q_++q_-+q_3$). 

Define $\mathbb Z_+ := \mathbb N_0$, $\mathbb Z_- :=\mathbb Z\setminus \mathbb N_0$,
and $F_{\alpha,\qq}:=x_\alpha\cdot F_{\qq}$, with $\alpha=\pm$).
Setting $\lambda:={\rm e}^{\mu\Gamma}$ and plugging  \eqref{eq:2.10}
into \eqref{eq:2.11}, we find for $n=1$
\begin{subequations} \label{eq:2.12}
\begin{align}
\xi^{(1)}(\varphi,t) & = \int_{0}^{t} {\rm d}\tau \,
{\rm e}^{\mu\Gamma (\varphi,t,\tau)}\gamma_{0}(\varphi,\tau),
\label{eq:2.12a} \\
U^{(1)}(\varphi) & =\sum_{m\in\mathbb Z_-} \lambda^{-m-1}\int_{0}^{2\pi} {\rm d}\tau \,
{\rm e}^{\mu\Gamma (S^ {m}\varphi,2\pi,\tau)}\gamma_{0}(S^ {m}\varphi,\tau),
\label{eq:2.12b} \\
a_\alpha^{(1)}(\varphi,t) & = \int_{0}^{t} {\rm d} \tau \, F_{\alpha,{\bf 0}}(\varphi,\tau),
\label{eq:2.12c} \\
h_{\alpha}^{(1)}(\varphi) & = -\alpha \sum_{m\in\mathbb{Z}_{\alpha}} 
\lambda_{\alpha}^{-m-1}\int_{0}^{2\pi} {\rm d} \tau \, 
F_{\alpha,{\bf 0}}(S^ {m}\varphi,\tau) .
\label{eq:2.12d} 
\end{align}
\end{subequations}

Introduce the notation 
\begin{eqnarray} \label{eq:2.12bis}
& & P_{\qq,\underline n}(\f,\tau) := \Biggl( \prod_{i=1}^{q_+} \bigl( \lambda_{+} h_{+}^{(n_{i})}
(\varphi) + a_{+}^{(n_{i})}(\varphi,\tau) \bigr) \Biggr)\cdot \\
& & \cdot\Biggl( \prod_{i=q_++1}^{q_++q_-} \bigl( \lambda_- h_-^{(n_{i})}
(\varphi) + a_-^{(n_{i})}(\varphi,\tau) \bigr) \Biggr)
\Biggl( \prod_{i=q_++q_-+1}^{|\qq|} \bigl( {\rm e}^{\mu\Gamma(\varphi,\tau,0)}
U^{(n_{i})}(\varphi)+\xi^{(n_{i})}(\varphi,\tau) \bigr) \Biggr) , \nonumber
\end{eqnarray}
where $\underline{n}:=(n_{1},\ldots,n_{|\qq|})$. Here and henceforth,
if $q_+=0$, the product $\prod_{i=1}^{q_+}(\l_+h_+^{(n_i)}(\f)+a_+^{(n_i)}(\f,\tau))$
should be interpreted as 1, and similarly for the other products in the case that $q_-=0$ and/or $q_3=0$.
Then, defining $\mathbb  N_1:=\mathbb Z\cap(0,+\infty)$, we find, for $n\geq2$, 
\begin{subequations} \label{eq:2.13}
\begin{align}
\xi^{(n)}(\varphi,t) & = \int_{0}^{t} {\rm d}\tau \,
{\rm e}^{\mu\Gamma(\varphi,t,\tau)}
\sideset{}{^*}\sum_{\qq\in\mathbb N_0^3} G_{\qq}(\varphi,\tau)
\sum_{\substack{\underline n\in\mathbb N_1^{|\qq|}:\\ |\underline n|=n-1}}
P_{\qq,\underline n}(\f,\tau) ,
\label{eq:2.13a} \\
U^{(n)}(\varphi) & = \sum_{m\in\mathbb Z_-}\l^{-m-1} \int_{0}^{2\p} {\rm d}\tau \,
{\rm e}^{\mu\Gamma(S^m\varphi,2\p,\tau)}
\sideset{}{^*}\sum_{\qq\in\mathbb N_0^3} G_{\qq}(S^m\varphi,\tau)
\sum_{\substack{\underline n\in\mathbb N_1^{|\qq|}:\\ |\underline n|=n-1}}
P_{\qq,\underline n}(S^m\f,\tau) ,
\label{eq:2.13b} \\
a_{\alpha}^{(n)}(\varphi,t) & =
\int_{0}^{t} {\rm d}\tau \,\sum_{\qq\in\mathbb N_0^3}  F_{\a,\qq}(\varphi,\tau)
\sum_{\substack{\underline n\in\mathbb N_1^{|\qq|}:\\ |\underline n|=n-1}}
P_{\qq,\underline n}(\f,\tau) ,
\label{eq:2.13c} \\
h_{\alpha}^{(n)}(\varphi) & = - \alpha \sum_{m\in\mathbb{Z}_{\alpha}}
\lambda_{\alpha}^{-m-1} \int_{0}^{2\p} {\rm d}\tau \,
\sum_{\qq\in\mathbb N_0^3} F_{\a,\qq}(S^m\varphi,\tau)
\sum_{\substack{\underline n\in\mathbb N_1^{|\qq|}:\\ |\underline n|=n-1}}
P_{\qq,\underline n}(S^m\f,\tau) ,
\label{eq:2.13d} 
\end{align}
\end{subequations}
where $|\underline{n}|:=n_{1}+\ldots+n_{|\qq|}$.

\subsection{Tree expansion and convergence}

We now want to bound the generic term in the series originating from the recursive equations
\eqref{eq:2.13}; the goal is to show that the  $n$-th order
is bounded proportionally to  $\left(C\m^{-\a}\right)^{n}$, with $C>0$ and $0<\a<1$. 
We find convenient to represent graphically the coefficients in \eqref{eq:2.10}
in terms of rooted trees (or simply trees, in the following) as in \cite{GGG}. 
We refer to \cite[Section V]{GGG} for the definition of trees and notations.
With respect to the trees in \cite{GGG} in the present case there are four {\it types} of nodes.
We use the symbols 
\raisebox{1mm}{\xymatrix{*+[o][F*:black]{}}},
\raisebox{1mm}{\xymatrix{*+[o][F]{}}},
\raisebox{1mm}{\xymatrix{*+[F*:black]{}}} and 
\raisebox{1mm}{\xymatrix{*+[F]{}}}, calling them nodes of type
$0$, $1$, $2$ and $3$, respectively: they correspond to contributions to $\xi,U,a,h$, respectively. 
The constraint $*$ forbids a node of type 0 or 1 to be immediately preceded by exactly one node of these two types.

Recall that a \emph{tree} is a partially ordered set of nodes and lines; the partial ordering relation
is denoted by $\succeq$ and each line will be drawn as an arrow pointing from the node it exits to the node it enters.
We call $V(\theta)$ the set of nodes and $L(\theta)$ the set of lines of the tree $\theta$. 
As in \cite{GGG} we denote by $v_{0}$ the node such that
$v_{0}  \succeq v$ for any node $v \in V(\theta)$: $v_{0}$ will be called the \emph{special node}
and the line exiting $v_{0}$ will be called the \emph{root line}. The root line can be imagined to enter
a further point, called the root, which, however, is not counted as a node.

With each node we associate a label $\eta_{v}\in\{0,1,2,3\}$ to denote its type
and a time variable $\tau_{v}\in(0,2\pi]$. With the nodes of types 2 and 3 we also associate a label 
$\alpha_{v}\in\{\pm\}$.
Denoting by $\mathcal{P}(v,w)$ the path of lines connecting $w$ to $v\prec w$, 
with both $v$ and $w$ included, set
\begin{equation} \nonumber
m(v) = \sum_{w \in \mathcal{P}(v,v_0)} m_{w} ,\end{equation}
where $m_{v}\in\mathbb Z_-$ if $\eta_{v}=1$, $m_v\in\mathbb Z_{\a_v}$ if $\eta_v=3$, and $m_v=0$ otherwise.
Given a node $v$, we denote by $v'$ the unique node immediately following it; moreover, we let $s_{v}^{\eta}$
be the number of nodes of type $\eta$ immediately preceding it; if $\eta=2,3$, we also define $s_v^{\eta,\a}$, $\a=\pm$,
to be the number of nodes $w$ of type $\eta$ with $\a_{w}=\a$ immediately preceding it (i.e. such that $w'=v$).
Finally, we let $\qq_v=(q_{v,+},q_{v,-},q_{v,3})$, with $q_{v,\pm}=s^{2,\pm}_v+s^{3,\pm}_v$, 
and $q_{v,3}=s^0_v+s^1_v$. A node $v$ is called an \emph{end-node} if $\qq_v=(0,0,0)$, while it is called
an \emph{internal node} if it is not an end-node.

The \emph{node factor} $A_{v}=A_{v}(\varphi,\tau_{v'},\tau_{v})$ is defined as
\begin{equation} \nonumber
A_{v} := \begin{cases}
\displaystyle{{\rm e}^{\mu\Gamma(S^ {m(v)}\varphi,\tau_{v'},\tau_{v})}
G_{\qq_v}(S^ {m(v)}\varphi,\tau_{v})
\, {\rm e}^{\mu s_{v}^{1} \Gamma(S^{m(v)}\varphi,\tau_{v},0)} \l_+^{q_{v,+}} \l_-^{q_{v,-}}}, &
\qquad {\rm if}\ \eta_{v} = 0 , \\
\displaystyle{\lambda^{-m_{v}-1}
{\rm e}^{\mu\Gamma(S^ {m(v)}\varphi,2\pi,\tau_{v})}
G_{\qq_v}(S^ {m(v)}\varphi,\tau_{v})
\, {\rm e}^{\mu s_{v}^{1} \Gamma(S^{m(v)}\varphi,\tau_{v},0)}  \l_+^{q_{v,+}} \l_-^{q_{v,-}}}, &
\qquad {\rm if}\ \eta_{v} = 1 , \\
\displaystyle{
F_{\a_v,\qq_{v}}(S^ {m(v)}\varphi,\tau_{v})
\, {\rm e}^{\mu s_{v}^{1} \Gamma(S^{m(v)}\varphi,\tau_{v},0)}  \l_+^{q_{v,+}} \l_-^{q_{v,-}}}, &
\qquad {\rm if}\ \eta_{v} = 2 , \\
\displaystyle{-\alpha_{v} \lambda_{\alpha_{v}}^{-m_{v}-1}
F_{\a_v,\qq_v}(S^ {m(v)}\varphi,\tau_{v})
\, {\rm e}^{\mu s_{v}^{1} \Gamma(S^{m(v)}\varphi,\tau_{v},0)}  \l_+^{q_{+,v}} \l_-^{q_{-,v}}}, &
\qquad {\rm if}\ \eta_{v} = 3 ,
\end{cases}
\end{equation}
where $\tau_{v_{0}'}$ is interpreted as equal to $t$, while the \emph{node integral} is 
\begin{equation} \nonumber
I_{v} := \begin{cases}
\displaystyle{\int_{0}^{\tau_{v'}} {\rm d}\tau_{v}} \, & \qquad {\rm if}\ \eta_{v}=0,2 , \\
\displaystyle{\int_{0}^{2\pi} {\rm d}\tau_{v}} \, & \qquad {\rm if}\  \eta_{v}=1,3 .
\end{cases}
\end{equation}
With the definitions above, we denote by $\Theta_{n,\eta}^{*}$ the set of labelled trees with $n$ nodes, 
$\eta_{v_0}=\eta$, and the constraint that nodes of type $0$ or $1$ cannot be immediately preceded by exactly 
one node of type $0$ or $1$; if $\eta=2,3$, we also denote  by $\Theta_{n,\eta,\a}^{*}$
the subset of  by $\Theta_{n,\eta}^{*}$ with $\a_{v_0}=\a$. Then, one can prove by induction that 
\begin{subequations} \label{eq:2.14}
\begin{align}
& \xi^{(n)}(\varphi,t) = \sum_{\theta\in\Theta^{*}_{n,0}} {\rm Val}(\theta) , \qquad
U^{(n)}(\varphi) = \sum_{\theta\in\Theta^{*}_{n,1}} {\rm Val}(\theta) ,
\label{eq:2.14a} \\
& a^{(n)}_\a(\varphi,t) = \sum_{\theta\in\Theta^{*}_{n,2,\a}} {\rm Val}(\theta) , \qquad
h_\a^{(n)}(\varphi) = \sum_{\theta\in\Theta^{*}_{n,3,\a}} {\rm Val}(\theta) .
\label{eq:2.14b}
\end{align}
\end{subequations}
where
\begin{equation} \nonumber
{\rm Val} {(\theta)} = \prod_{v\in V(\theta)} I_{v} A_{v} ,
\end{equation}
with the integrals to be performed by following the tree ordering,
i.e. by starting from the end-nodes and by moving towards the root. 

In Figure \ref{fig:n=00003D1} the first order contributions are graphically represented, while the contributions  of order $n=2$
are shown in Figure \ref{fig:n=00003D2}. For each node $v$, the label $\a_v$ is drawn superimposed
on the line exiting $v$, for clarity purposes, while the label $m_{v}$ (to be summed over) is not explicitly shown.

\begin{figure}[h]
\centering
\begin{raggedright}
\par\end{raggedright}
\begin{raggedright}
\xymatrix@C=2cm@R=0.5cm{
\hspace{2.8cm}
{\xi^{(1)}(\varphi,t)=}\ar@{<-}[r]^-{} _(1.08){v_0} &*+[o][F*:black]{} & 
{U^{(1)}(\varphi)=}\ar@{<-}[r]^-{} _(1.10){v_0} &*+[o][F]{} \\
\hspace{2.8cm}
{a^{(1)}_{\alpha}(\varphi,t)=}\ar@{<-}[r]^-{\alpha} _(1.08){v_0} &*+[F*:black]{} & 
{h^{(1)}_{\alpha}(\varphi)=}\ar@{<-}[r]^-{\alpha} _(1.12){v_0} &*+[F]{} }  
\par\end{raggedright}
\raggedright{}\caption{\label{fig:n=00003D1}\small Graphical representation of $\xi^{(1)}(\varphi,t)$,
$U^{(1)}(\varphi)$, $a_{\alpha}^{(1)}(\varphi,t)$ and $h_{\alpha}^{(1)}(\varphi)$.}
\end{figure}

\begin{figure}[h]
\centering
\begin{raggedright}
\par\end{raggedright}
\xymatrix@C=1.5cm@R=0.5cm{{
\hskip.1cm 
\hspace{1.6cm}
\xi^{(2)}(\varphi)=}\ar@{<-}[r]^-{} _(1.12){v_0} & *+[o][F*:black]{}\ar@{<-}[r]^-{\alpha'} _(1.20){v_1} & *+[F]{} & +
& {}\ar@{<-}[r]^-{} _(1.20){v_0} & *+[o][F*:black]{}\ar@{<-}[r]^-{\alpha'} _(1.20){v_1} & *+[F*:black]{} }
\vspace{0.8cm}
\xymatrix@C=1.5cm@R=0.5cm{{
\hspace{1.6cm}
U^{(2)}(\varphi)=}\ar@{<-}[r]^-{} _(1.12){v_0} & *+[o][F]{}\ar@{<-}[r]^-{\alpha'} _(1.20){v_1} & *+[F]{} & +
& {}\ar@{<-}[r]^-{} _(1.20){v_0} & *+[o][F]{}\ar@{<-}[r]^-{\alpha'} _(1.20){v_1} & *+[F*:black]{} }
\vspace{0.8cm}
\xymatrix@C=1.5cm@R=0.5cm{{
\hspace{1.6cm}
a^{(2)}_{\alpha}(\varphi)=}\ar@{<-}[r]^-{\alpha} _(1.12){v_0}
& *+[F*:black]{}\ar@{<-}[r]^-{\alpha'} _(1.20){v_1} & *+[F]{} & +
& {}\ar@{<-}[r]^-{\alpha} _(1.20){v_0} & *+[F*:black]{}\ar@{<-}[r]^-{\alpha'} _(1.20){v_1} & *+[F*:black]{} \\ 
\hspace{1.6cm}
{\hspace{0.5cm}+\hspace{0.8cm}} 
\ar@{<-}[r]^-{\alpha} _(1.14){v_0} & *+[F*:black]{}\ar@{<-}[r]^-{} _(1.20){v_1}
& *+[o][F]{} & + & {}\ar@{<-}[r]^ -{\alpha} _(1.20){v_0} & *+[F*:black]{}\ar@{<-}[r]^-{} _(1.20){v_1} & *+[o][F*:black]{} }
\vspace{0.8cm}
\xymatrix@C=1.5cm@R=0.5cm{{
\hspace{1.6cm}
h^{(2)}_{\alpha}(\varphi)=}\ar@{<-}[r]^-{\alpha}  _(1.16){v_0} & *+[F]{}\ar@{<-}[r]^-{\alpha'} _(1.20){v_1} & *+[F]{} & +
& {}\ar@{<-}[r]^-{\alpha} _(1.20){v_0} & *+[F]{}\ar@{<-}[r]^-{\alpha'} _(1.20){v_1} & *+[F*:black]{} \\ 
\hspace{1.6cm}
{\hspace{0.5cm}+\hspace{1.0cm}} 
\ar@{<-}[r]^-{\alpha} _(1.14){v_0} & *+[F]{}\ar@{<-}[r]^-{} _(1.20){v_1} 
& *+[o][F]{} & + & {}\ar@{<-}[r]^ -{\alpha} _(1.20){v_0} & *+[F]{}\ar@{<-}[r]^-{} _(1.20){v_1} & *+[o][F*:black]{} }
\caption{\label{fig:n=00003D2}\small Graphical representation of $\xi^{(2)}(\varphi,t)$, $U^{(2)}(\varphi,t)$,
$a_{\alpha}^{(2)}(\varphi,t)$ and $h_{\alpha}^{(2)}(\varphi,t)$. }
\end{figure}

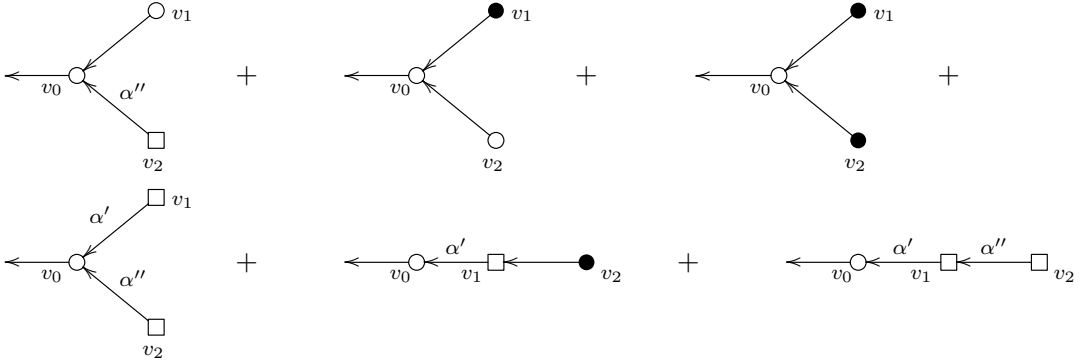
\begin{figure}[h]
\vspace{1cm}
\centering
\begin{raggedright}
\par\end{raggedright}
\xymatrix@R=0.5cm{ &  & *+[o][F]{} & & &  & *+[o][F*:black]{} & & &  & *+[o][F*:black]{} \\ 
\hspace{1.2cm}
{}\ar@{<-}[r] _(0.80){v_0} & *+[o][F]{}\ar@{<-}[ru] _(1.14){v_1} \ar@{<-}[rd]^-{\alpha''} _(1.16){v_2} &  & + 
& {}\ar@{<-}[r] _(0.80){v_0} & *+[o][F]{}\ar@{<-}[ru] _(1.14){v_1} \ar@{<-}[rd] _(1.16){v_2} & & +
& {}\ar@{<-}[r] _(0.80){v_0} & *+[o][F]{}\ar@{<-}[ru] _(1.14){v_1} \ar@{<-}[rd] _(1.16){v_2} &  & + \\
&  & *+[F]{} & & &  & *+[o][F]{} & & &  & *+[o][F*:black]{} & \\ 
&  & *+[F]{}  \\ 
\hspace{1.2cm}
{}\ar@{<-}[r] _(0.80){v_0} & *+[o][F]{}\ar@{<-}[ru]^-{\alpha'} _(1.14){v_1} \ar@{<-}[rd]^-{\alpha''} _(1.16){v_2} & & +  
&  {}\ar@{<-}[r] _(0.80){v_0} & *+[o][F]{}\ar@{<-}[r]^-{\alpha'} _(0.70){v_1} & *+[F]{}\ar@{<-}[r]^-{} _(1.28){v_2} & *+[o][F*:black]{}  &  + 
& {}\ar@{<-}[r] _(0.80){v_0} & *+[o][F]{}\ar@{<-}[r]^{\alpha'} _(0.70){v_1} & *+[F]{}\ar@{<-}[r]^-{\alpha''} _(1.28){v_2} & *+[F]{} \\  & & *+[F]{}               }

\caption{\label{fig:n=00003D3}\small Graphical representation of some contributions to $U^{(3)}(\varphi,t)$.}
\end{figure}

For $n=3$ a few contributions to $U^{(3)}(\varphi)$ are shown in Figure \ref{fig:n=00003D3}. All the other contributions are
obtained by replacing the nodes preceding the special node $v_{0}$ 
by nodes of a different type, with the constraint that if only one line enters $v_{0}$ then
it exits a node of type 2 or 3; note that in \cite{GGG} the linear trees at the bottom of the figure were not possible. 

Given $\bar\theta\in  \Theta^*_{n,\eta}$, we let $\mathcal{T}(\bar\theta)$ be the family of labelled trees differing from 
$\bar\theta$ just by the choice of the labels $\{m_v\}$. Then, 
using \eqref{2.an} and proceeding as in \cite[Section VI]{GGG}, we obtain 
\begin{equation} \label{eq:2.15}\sum_{\theta\in\mathcal{T}(\bar\theta)}
\left| {\rm Val}(\theta) \right| \le C_{1}^{n} \mu^{-N_{i}^{1}(\bar\theta)}\;,
\end{equation}
where $C_1>0$ is a suitable constant and $N_{i}^{\eta}(\bar\theta)$ is the number of internal nodes of type $\eta$
in $\bar\theta$; see \cite{GGG} for details. We are then left with bounding $N_i^1(\bar\theta)$.

\begin{lemma} \label{lem:2.1}
For all $\theta\in\Theta_{n,\eta}^{*}$ one has $\displaystyle{N_{i}^{1}(\theta)\leq\frac{2n-1}{3}}$.
\end{lemma}

The proof is given in Appendix \ref{app:a}, where it is also shown that such an upper bound on 
$N_i^1(\theta)$ is optimal, i.e. there are trees that saturate the inequality. Combining Lemma \ref{lem:2.1} 
with \eqref{eq:2.15} and recalling that the number of distinct families $\mathcal{T}(\bar\theta)$
in $\Theta^*_{n,\eta}$ is bounded by $C_2^n$, for a suitable $C_2$, we get
\begin{equation} \label{eq:2.16}
\max \left\{ |\xi^{(n)}(\varphi,t)|,|U^{(n)}(\varphi) |,
|a_\a^{(n)}(\varphi,t)|, |h_\a^{(n)}(\varphi)| \right\} \leq C_{3}^{n}\mu^{-[(2n-1)/3]} ,
\end{equation}
where $C_{3}$ is a positive constant and $[\cdot]$ is the integer part. Eq.\eqref{eq:2.16} implies that the radius of convergence of the series \eqref{eq:2.10} is bounded by $\varepsilon_{0}=C_{3}^{-1}\mu^{2/3}$. Therefore, we can take $\mu=\varepsilon$. 
H\"older-continuity of $U$ and $H$ can be proved mutatis mutandis like in \cite{GGG}.
This completes the proof of Theorem \ref{prop:1}.

\section{Lyapunov Exponents} \label{sec:3}

\subsection{Conjugation of the tangent dynamics} \label{sec:3.1}

In order to compute the Lyapunov exponents, we need to understand how the vectors on the tangent space evolve under the 
interacting dynamics. To this purpose, we set $\boldsymbol{x}:=(x_{1},x_{2},x_{3})=(x_{1},x_{2},w)=(x,w) \in \mathbb{T}^{3}$, 
rewrite \eqref{eq:1.2} as $\dot{\boldsymbol{x}}=\boldsymbol{f}_0(\boldsymbol{x},t)
+\varepsilon\, \boldsymbol{f}(\boldsymbol{x},t)$, with $\boldsymbol{f}=(f_{1},f_{2},f_{3}):=(f_{1},f_{2},g)$,
and write the dynamics on the tangent space as follows: 
\begin{equation} \label{eq:3.1}
\dot{\boldsymbol{y}} =
A_t(S^{t}_{\varepsilon}(\boldsymbol{x})) \, \boldsymbol{y} , \qquad
A_t(\boldsymbol{x})=\delta(t) \big[P_+\log\l_++P_-\log\l_-\big] 
+\e \boldsymbol{\partial f}({\boldsymbol{x}},t)\;,
\end{equation}
where $\boldsymbol{y}\in\mathbb{R}^{3}$, $S^{t}_{\varepsilon}(\boldsymbol{x})=(x(t),w(t))$ is the solution
\eqref{eq:2.9} to \eqref{eq:1.2} found in Section \ref{sec:2}, $P_\pm$ are the projections
into the unperturbed eigendirections $(x_\pm,0)$, $\l_\pm$ are the corresponding unperturbed eigenvalues, 
and $\boldsymbol{\partial f}$ is the Jacobian matrix of $\boldsymbol{f}$.

Integration of \eqref{eq:3.1} gives the tangent map $DS_{\varepsilon}^{t}:T_{\boldsymbol{x}}\mathbb{T}^{3} = 
\mathbb{R}^{3} \to T_{S_{\varepsilon}^{t}(\boldsymbol{x})}\mathbb{T}^{3} =\mathbb{R}^{3}$. We denote by
$DS^t_\e(\boldsymbol{x})\boldsymbol{y}$ the solution to \eqref{eq:3.1}
with initial condition $\boldsymbol{y}$ at $t=0$ (started at $S^0_\e(\boldsymbol{x})=\boldsymbol{x}$). 
For $t\in (0,2\pi]$ we obtain
\begin{equation} \label{eq:3.2}
DS_{\varepsilon}^{t}(\boldsymbol{x})\, \boldsymbol{y}=
\L \boldsymbol{y}+\varepsilon\int_{0}^{t}
{\rm d}\tau \, \boldsymbol{\partial f}(S^{\tau}_{\varepsilon}(\boldsymbol{x}),\tau)
DS^{\tau}_{\varepsilon}(\boldsymbol{x})\,\boldsymbol{y} ,
\qquad
\L=\left( \begin{matrix} 2 & 1 & 0\\ 1 & 1 & 0\\ 0 & 0 & 1 \end{matrix} \right) .
\end{equation}

We look for a conjugation $\mathcal{H}:\mathbb{T}^{2}\times\mathbb{R}^{3} \to \mathbb{T}^{3}\times\mathbb{R}^{3}$,
\begin{equation} \label{eq:3.3}
\mathcal{H}:(\varphi, \boldsymbol{Y})\longrightarrow
(\boldsymbol{H}(\varphi),(\mathds{1}+K(\varphi)) \boldsymbol{Y}) , \qquad
\boldsymbol{H}(\varphi)=(H(\varphi),W(\varphi)) ,
\end{equation}
where $\mathds{1}$ is the identity in $\mathbb{R}^{3}$ and $K(\f)$ is a $3\times3$ matrix, such that, by setting
\begin{equation} \nonumber
\Phi^{t}_{\varepsilon}(\boldsymbol{x},\boldsymbol{y}) =
(S^{t}_{\varepsilon} (\boldsymbol{x}),DS^{t}_{\varepsilon}(\boldsymbol{x}) \boldsymbol{y}) , \qquad
\widehat{S}_{0,\varepsilon}
(\varphi,\boldsymbol{Y})=(S  \varphi, (\L+N(\varphi)) \, \boldsymbol{Y})
\end{equation}
for a suitable $3\times3$ matrix $N(\f)$, to be determined, one has $\Phi^{2\pi}_{\varepsilon} \circ \mathcal{H} = \mathcal{H} \circ \widehat{S}_{0,\varepsilon}$, that is
\begin{equation} \label{eq:3.4}
\Phi_{\varepsilon}^{2\pi}(\boldsymbol{H}(\varphi),(\mathds{1}+K(\varphi)) \, \boldsymbol{Y}) =
(\boldsymbol{H}(S \varphi),(\mathds{1}+K(S \varphi)) \, (\L+ N(\varphi)) \, \boldsymbol{Y}) .
\end{equation}
Of course only the part of \eqref{eq:3.4} involving the tangent dynamics has still to be solved, so we  
study the conjugation equation
\begin{equation} \label{eq:3.5}
DS_{\varepsilon}^{2\pi}(\boldsymbol{H}(\varphi)) \, (\mathds{1}+K(\varphi)) \, \boldsymbol{Y}
= (\mathds{1}+K(S \varphi)) \, (\L+N(\varphi)) \, \boldsymbol{Y} .
\end{equation}

The matrix $N(\varphi)$ will be taken to be diagonal in the basis
$\left\{ \boldsymbol{y}_{+},\boldsymbol{y}_{-},\boldsymbol{y}_{3}\right\}$,
where $\boldsymbol{y}_{\pm}=(x_{\pm},0)$, and $\boldsymbol{y}_{3}=(0,0,1)$.
Then in the basis $\left\{ \boldsymbol{y}_{+},\boldsymbol{y}_{-},\boldsymbol{y}_{3}\right\} $ one has
\begin{equation} \nonumber
K(\varphi) = \left(\begin{matrix}
k_{++}(\varphi) & k_{+-}(\varphi) & k_{+3}(\varphi)\\
k_{-+}(\varphi) & k_{--}(\varphi) & k_{-3}(\varphi)\\
k_{3+}(\mathbf{\varphi}) & k_{3-}(\varphi) & k_{33}(\varphi) \end{matrix} \right) , \qquad
N(\varphi)=\left(\begin{matrix}
\nu_{+}(\varphi) & 0 & 0\\
0 & \nu_{-}(\varphi) & 0\\
0 & 0 & \nu_3(\varphi)
\end{matrix} \right) ,
\end{equation}
while the matrix $\L$ takes diagonal form, with values $\boldsymbol{\l}=(\l_+,\l_-,1):=(\l_+,\l_-,\l_3)$ along the main diagonal.
From now on we shall use this basis, and implicitly assume that the indices $i,j,\ldots$, run over the values $+,-,3$,
unless stated otherwise.

By setting $\boldsymbol{y}_{i}(\varphi):=(\mathds{1}+K(\varphi))\boldsymbol{y}_{i}$, we obtain from \eqref{eq:3.5}
\begin{equation} \label{eq:3.6}
DS_{\varepsilon}^{2\pi}(H(\varphi),W(\varphi)) \, \boldsymbol{y}_{i}(\varphi) =
\lambda_{i}(\varphi) \, \boldsymbol{y}_{i}(S \varphi), \qquad
\lambda_{i}(\varphi) := \lambda_{i}+\nu_{i}(\varphi).
\end{equation}
Note that, given a solution $y_i(\f)$ of \eqref{eq:3.6}, then also $\boldsymbol{y}_{i}'(\varphi):=
l_{i}(\varphi)\boldsymbol{y}_{i}(\varphi)$ is a solution with $\lambda_{i}(\f)$ replaced with $\lambda_{i}'(\varphi)
=(l_{i}(\varphi)/l_{i}(S\varphi))\,\lambda_{i}(\varphi)$, where $l_{i}(\varphi)$ are non-zero functions from $\mathbb{T}^{2}$
to $\mathbb{R}$. Therefore, with no loss of generality, we can require the diagonal elements of $K(\varphi)$ to vanish:
hence $K(\varphi)$ will be looked for  as an off-diagonal matrix.

\subsection{Recursive equations} \label{sec:3.2}

We look for the conjugation in the form \eqref{eq:3.3}-\eqref{eq:3.4}, with
\begin{equation} \label{eq:3.7}
k_{ij}(\varphi) = \sum_{n=1}^{\infty} \varepsilon^{n} k_{ij}^{(n)}(\varphi) , \qquad
\nu_{i}(\varphi) = \sum_{n=1}^{\infty} \varepsilon^{n} \nu_{i}^{(n)}(\varphi) .
\end{equation}

Equations \eqref{eq:3.2} and \eqref{eq:3.5} give
\begin{eqnarray} 
& & \hskip-.5truecm
\L \, (\mathds{1}+K(\varphi)) + \varepsilon
\int_{0}^{2\pi} {\rm d}\tau \, \boldsymbol{\partial f}(S^{\tau}_{\varepsilon}(\boldsymbol{H}(\varphi)),\tau) \,
DS_{\varepsilon}^{\tau} (\boldsymbol{H}(\varphi)) \, (\mathds{1}+K(\varphi)) \nonumber \\
& & \hskip-.5truecm
\qquad \qquad  = \L + N(\varphi) + K(S \varphi) \L + K(S \varphi) N(\varphi) , 
\label{eq:3.8}
\end{eqnarray}
where $DS_{\varepsilon}^{\tau}(\boldsymbol{H}(\varphi))$ can be computed iteratively via \eqref{eq:3.2} as
\bea &&\hskip-.8truecm DS_{\varepsilon}^{\tau}(\boldsymbol{H}(\varphi))= \L+\sum_{n\ge 1}\e^n\int_0^\t \!\!\!d\t_1
\boldsymbol{\partial f}(S^{\tau_1}_{\varepsilon}(\boldsymbol{H}(\varphi)),\tau_1)
\int_0^{\t_1}\!\!\!d\t_2\boldsymbol{\partial f}(S^{\tau_2}_{\varepsilon}(\boldsymbol{H}(\varphi)),\tau_2)\cdots 
\nonumber\\
&& \hskip1.15truecm \cdots 
\int_0^{\t_{n-1}}\!\!\!\!\!d\t_n\boldsymbol{\partial f}(S^{\tau_n}_{\varepsilon}(\boldsymbol{H}(\varphi)),\tau_n)\L
:=(\mathds{1}+\mathfrak M(\f,\t))\L.\label{eq:3.9bis}\eea
At first order in $\e$, \eqref{eq:3.8} gives (with $i,j\in\{\pm,3\}$)
\begin{equation} \label{eq:3.10}\l_{i}k_{ij}^{(1)}(\varphi)-\l_{j}k_{ij}^{(1)}(S\varphi)  + \l_{j}
\int_{0}^{2\pi} {\rm d}\tau \, \partial_{j} f_{i} (S \varphi,w_0+\tau,\tau) =\nu^{(1)}_{i}(\varphi) \, \delta_{ij}.
\end{equation}
In particular, setting $i=j$ and recalling that $k^{(1)}$ is off-diagonal, we find
\begin{equation} \nonumber
\nu_{\pm}^{(1)}(\varphi) =\l_\pm \int_{0}^{2\pi} {\rm d} \tau \,
\partial_{\pm}f_{\pm}(S\varphi,w_{0}+\tau,\tau) , \qquad
\nu_{3}^{(1)}(\varphi) = \int_{0}^{2\pi} {\rm d} \tau \,
\partial_{w} g(S\varphi,w_{0}+\tau,\tau)\equiv\Gamma ,
\end{equation}
while, for $i\neq j$, we have to solve recursively \eqref{eq:3.10} for $k_{ij}^{(1)}(\varphi)$, the result being (if $\a=\pm$): 
\begin{subequations} \label{eq:3.11}
\begin{align}
k_{\alpha,-\a}^{(1)}(\varphi) & =-\a\sum_{m\in\mathbb Z_\a} \left( \frac{\l_\a}{\l_{-\a}} \right)^{-m-1}
\int_{0}^{2\pi} {\rm d} \tau \,
\partial_{-\a}f_{\alpha}(S^{m+1}\varphi,w_{0}+\tau,\tau),
\label{eq:3.11a} \\
k_{\alpha,3}^{(1)}(\varphi) & =-\a \sum_{m\in\mathbb{Z_{\alpha}}} \lambda_{\a}^{-m-1}
\int_{0}^{2\pi} {\rm d} \tau \, \partial_{w}f_{\alpha}(S^ {m+1}\varphi,w_{0}+\tau,\tau),
\label{eq:3.11b} \\
k_{3,-\alpha}^{(1)}(\varphi) & = -\alpha \sum_{m\in\mathbb{Z}_{\a}} \lambda_{-\a}^{m+1}
\int_{0}^{2\pi} {\rm d} \tau \, \partial_{-\alpha}f_{3}(S^ {m+1}\varphi,w_{0}+\tau, \tau) .
\label{eq:3.11c}
\end{align}
\end{subequations}
In order to compute the higher orders, we insert \eqref{eq:3.9bis} in the left side of \eqref{eq:3.8}, thus getting
\begin{eqnarray}
& & \Lambda K(\varphi) - K(S \varphi)\,\Lambda - N(\varphi) - 
K(S \varphi)N(\varphi) +\mathfrak{M}(\f)\L(\mathds{1}+K(\varphi)) = 0 ,
\label{3.11d}
\end{eqnarray}
where $\mathfrak{M}(\f):=\mathfrak{M}(\f,2\p)$. Note that, according to \eqref{eq:3.9bis},
$\mathfrak{M}(\f)$ is expressed as a series
of iterated integrals of $\boldsymbol{\partial f}(S^{\tau}_{\varepsilon}(\boldsymbol{H}(\varphi)),\t)$,
where $\boldsymbol{\partial f}$ is analytic in its argument. Therefore, the power series expansion
in $\e$ of $\mathfrak{M}(\f)$ can be obtained (and its $n$-th order coefficient can be bounded) by using  
the corresponding expansions for 
the components of $S^{\tau}_{\varepsilon}(\boldsymbol{H}(\varphi))=
(S H(\varphi)+a(\varphi,\tau),w_{0}+\tau+{\rm e}^{\mu \Gamma(\varphi,\tau,0)}U(\varphi)+\xi(\varphi,t))$; here 
the functions $H(\varphi)$, $a(\varphi,\tau)$, $U(\varphi)$ and $\xi(\varphi,\tau)$
are as in \eqref{eq:2.10} with the coefficients given by \eqref{eq:2.14} and bounded as in \eqref{eq:2.16}. We write 
\begin{equation} \label{eq:3.13}
\mathfrak{M}(\varphi) = \sum_{n=1}^{\infty} \varepsilon^{n} \mathfrak{M}^{(n)}(\varphi). 
\end{equation}
By using the very definition of $\mathfrak{M}(\f)$ and the bounds \eqref{eq:2.16}, it is straightforward to prove that 
\be \| \mathfrak{M}^{(n)}(\f) \| := \max_{i,j \in\{+,-,3\}} \left| \mathfrak{M}_{i,j}(\f) \right| \le C_4^n\m^{-[(2n-1)/3]}\label{e3.14}\ee
for a suitable $C_4>0$. Now, if $n\ge 2$, the diagonal part of \eqref{3.11d} gives
\begin{equation} \label{eq:3.15}
\nu_{i}^{(n)}(\varphi) =\mathfrak{M}_{ii}^{(n)}(\f)\l_i+
\sum_{\substack{n_{1}+n_2=n\\ n_1,n_2\ge 1}}\ 
\sum_{j=\pm,3}\mathfrak{M}_{ij}^{(n_1)}(\f)\l_{j} 
k_{ji}^{(n_2)}(\f),
\end{equation}
while the off-diagonal part can be solved in a way similar to \eqref{eq:3.11}, i.e., if $\a=\pm$,
\begin{subequations} \label{eq:3.16}
\begin{align}
k_{\alpha,-\a}^{(n)}(\varphi) & = 
-\a\l_\a^{-1}\sum_{m\in\mathbb Z_\a} \left( \frac{\l_\a}{\l_{-\a}} \right)^{-m}
\Big[\mathfrak{M}_{\a,-\a}^{(n)}(S^m\f)\l_{-\a} + \mathcal Q_{\a,-\a}^{(n)}(S^m\f) \Bigr] ,
\label{eq:3.16a} \\
k_{\alpha,3}^{(n)}(\varphi) & = 
-\a\l_\a^{-1}\sum_{m\in\mathbb Z_\a}\l_\a^{-m}
\Big[ \mathfrak{M}_{\alpha,3}^{(n)}(S^m\f) + \mathcal Q_{\a,3}^{(n)}(S^m\f) \Bigr] ,
\label{eq:3.16b} \\
k_{3,-\alpha}^{(n)}(\varphi) & =- 
\a\sum_{m\in\mathbb Z_{\a}}\l_{-\a}^{m}
\Big[\mathfrak{M}_{3,-\alpha}^{(n)}(S^m\f)\l_{-\a} + \mathcal Q_{3,-\a}^{(n)}(S^m\f) \Bigr] ,
\label{eq:3.16c}
\end{align}
\end{subequations}
where we have set
\begin{equation} \label{eq:3.16d}
\mathcal Q_{i,j}^{(n)}(\f) := \sum_{\substack{n_{1}+n_2=n\\ n_1,n_2\ge 1}}
\Big(-k_{i,j}^{(n_1)}(S\f)\nu_{j}^{(n_2)}(\f)+\sum_{j'=\pm,3}\mathfrak{M}_{i,j'}^{(n_1)}(\f)\l_{j'} 
k_{j',j}^{(n_2)}(\f)\Big) .
\end{equation}
In the simple case that ${\boldsymbol f}=(0,0,g)$, $\nu_+(\f)=\nu_-(\f)=0$, while 
\begin{equation} 1+ \nu_3(\f)=1+\sum_{n\ge 1}\e^n\int_0^{2\pi}d\t_1\int_0^{\t_1} d\t_2\cdots\int_{0}^{\t_{n-1}}d\t \prod_{i=1}^n \partial_w g(S\f,w(\t_i),\t_i)=
e^{\e\int_0^{2\pi}d\t \partial_w g(S\f,w(\t),\t)},\nonumber
\end{equation}
thus recovering the formula for $A(2\pi)=\log(1+\nu_3(\f))$ given in \cite[Section VII]{GGG}. 

\subsection{Tree expansion and convergence} \label{sec:3.2}

In Figure \ref{fig:Rappresentazione-grafica-di Gamma^n} and \ref{fig:Rappresentazione-grafica-di K^n}
we give a graphical representation of \eqref{eq:3.15} and \eqref{eq:3.16a}, respectively. The representation of
\eqref{eq:3.16b} and \eqref{eq:3.16c} is the same as in Figure \ref{fig:Rappresentazione-grafica-di K^n},
simply with the labels $(\alpha,-\alpha)$ replaced by $(\alpha,3)$ and $(3,\alpha)$, respectively.

\begin{figure}[h]
\centering{}\xymatrix@R=0.5cm@C=1.0cm{ & & & & & & & &*+[o][F**:black]\txt{\hspace{0.2cm} } \\
\hspace{0.4cm}
\nu_i^{(n)}(\varphi)=\ar@{<-}[r] ^-{ii} ^(1.24){(n)} & *+[F**:black]\txt{\hspace{0.2cm} } 
& =\hspace{0.6cm}\ar@{<-}[r]^-{ii} _(0.75){v_0} & 
*+[F**:black]\txt{\hspace{0.2cm} }\ar@{<~}[r]^<<{
}^(0.6){ii} ^(1.4){(n)} 
&  *+[o][F**:black]\txt{\hspace{0.2cm} } & +
& 
\hspace{-.3cm} \displaystyle{\sum_{\substack{n_{1}+n_2=n\\ n_1,n_2\ge 1}}} \hspace{.5cm}
\ar@{<-}[r]^-{ii} _(0.80){v_0}
& *+[F**:black]\txt{\hspace{0.2cm} } \ar@{<~}[ru]^-{ij}_(1.2){(n_{1})}\ar@{<-}[rd]^(0.2){
}^-{ji}^(1.2){(n_{2})} &\\
& & & & & & & & *+[F**:white]\txt{\hspace{0.2cm} }     }\caption{
\small Graphical representation of $\nu_{i}^{(n)}(\varphi)$.\label{fig:Rappresentazione-grafica-di Gamma^n}}
\end{figure}

\begin{figure}[h]
\xymatrix@R=0.5cm@C=1.0cm{ & & & & & & & &*+[o][F**:black]\txt{\hspace{0.2cm} } \\
\hspace{0.4cm}
k_{i,j}^{(n)}(\varphi)=\ar@{<-}[r] ^-{i j} ^(1.24){(n)} & *+[F**:white]\txt{\hspace{0.2cm} } 
& =\hspace{0.6cm}\ar@{<-}[r]^-{i j} _(0.75){v_0} &
*+[F**:white]\txt{\hspace{0.2cm} }\ar@{<~}[r]^<<{
}^(0.7){i j} ^(1.4){(n)} & 
*+[o][F**:black]\txt{\hspace{0.2cm} } & + & 
\hspace{-0.3cm} {\displaystyle{\sum_{\substack{n_1,n_2\ge 1 \\ n_{1}+n_{2}=n}}}} \hspace{0.5cm}
\ar@{<-}[r]^-{i j} _(0.80){v_0} & 
*+[F**:white]\txt{\hspace{0.2cm} } \ar@{<~}[ru]^-{i j'}_(1.2){(n_{1})}
\ar@{<-}[rd]^(0.2){
}^(0.6){j' j}^(1.2){(n_{2})} &\\
& & & & & & & & *+[F**:white]\txt{\hspace{0.2cm} }\\  & & *+[F**:white]\txt{\hspace{0.2cm} } \\ 
\hspace{0.4cm} + \hspace{0.2cm}  {\displaystyle{\sum_{\substack{n_1,n_2\ge 1 \\ n_{1}+n_{2}=n}}}}
\ar@{<-}[r]^-{i j} _(0.80){v_0} & 
*+[F**:white]\txt{\hspace{0.2cm} } \ar@{<-}[ru]^-{i j}_(1.2){(n_1)}
\ar@{<-}[rd]^(0.25){\scriptscriptstyle{}}^(0.6){j j}^(1.2){(n_2)} & \\
& & *+[F**:black]\txt{\hspace{0.2cm} }     }
\caption{\small Graphical representation of $k_{i,j}^{(n)}(\varphi)$.\label{fig:Rappresentazione-grafica-di K^n}}
\end{figure}

To iterate the graphical construction and provide a tree representation for both $N(\varphi)$ and $K(\varphi)$,
we need a few more definitions. We identify three types of \emph{principal nodes}, that we call of type $K$, $N$
and $\mathfrak{M}$, and represent graphically, respectively, by
\raisebox{1mm}{\xymatrix{*+[F**:white]\txt{\hspace{0.2cm} }}} ,
\raisebox{1mm}{\xymatrix{*+[F**:black]\txt{\hspace{0.2cm} }}} and
\raisebox{1mm}{\xymatrix{*+[o][F**:black]\txt{\hspace{0.2cm} }}} .
With any such node $v$, we associate a label $\eta_{v}\in\{K,N,\mathfrak{M}\}$,
to denote its type, and two labels $i_{v},j_{v}\in\{+,-,3\}$,
which will be drawn superimposed to the line exiting $v$;
if $v$ is of type $N$, then $i_v=j_v$, while if $v$ is of type $K$, then $i_v\neq j_v$.
A node is of type $\mathfrak{M}$ if and only if it is an end-node. 
Furthermore, with each node $v$ with $\eta_{v}=\mathfrak{M}$, we associate a label $n_v\in \mathbb N_1$,
while we set $n_{v}=0$ for all nodes $v$ with $\eta_{v}\neq\mathfrak{M}$; 
with each node $v$ with $\eta_{v}=K$, we associate a label $\a_v\in\{+,-\}$ such that either $i_{v}=\alpha_v$ 
or $j_v=-\alpha_v$ (recall that if $v$ is a nodes of type $K$ then $i_v\neq j_v$, so that either $i_v$ or $j_v$ are $=\pm$), 
and a label $m_{v}\in\mathbb{Z}_{\alpha_v}$; if $v$ is of type $N$ or $\mathfrak{M}$ we define $m_v=0$. 

\begin{figure}[h]
\xymatrix{    & & *+[o][F**:black]\txt{\hspace{0.2cm} } &  & *+[o][F**:black]\txt{\hspace{0.2cm} }\\ 
\hspace{4.2cm} 
\ar@{<-}[r]^-{ij} _(0.88){v_0} &*+[F**:white]\txt{\hspace{0.2cm} }\ar@{<~}[ru]^-{i j'} _(0.90){v_1} \ar@{<-}[rd]^-{j' j} _(1.10){v_2} & 
&*+[F**:white]\txt{\hspace{0.2cm} } \ar@{<-}[rd]^-{j''' j} _(1.10){v_5} \ar@{<~}[ru]^-{j' j'''} _(0.90){v_4} & & *+[o][F**:black]\txt{\hspace{0.2cm} }  \\ 
& & *+[F**:white]\txt{\hspace{0.2cm} }\ar@{<-}[rrdd]^-{j j} _(1.10){v_6} \ar@{<-}[ru]_{j' j} _(0.86){v_3}
& &*+[F**:white]\txt{\hspace{0.2cm} } \ar@{<~}[ru]_-{j''' j} _(0.90){v_7} \\
& & & & & *+[o][F**:black]\txt{\hspace{0.2cm} }\\  
& & & & *+[F**:black]\txt{\hspace{0.2cm} } \ar@{<~}[ru]^-{j j''} _(0.90){v_8}
\ar@{<-}[rd]^-{j'' j} _(1.10){v_9} & & *+[o][F**:black]\txt{\hspace{0.2cm} } \\  
& & & & & *+[F**:white]\txt{\hspace{0.2cm} } \ar@{<~}[ru]_-{j'' j} _(0.90){v_{10}} \\            }
\caption{\label{fig:Esempi-di-alberi 1 e 2}A tree contributing to $k^{(n)}_{i,j}(\varphi)$ with $n=11$.}
\end{figure}
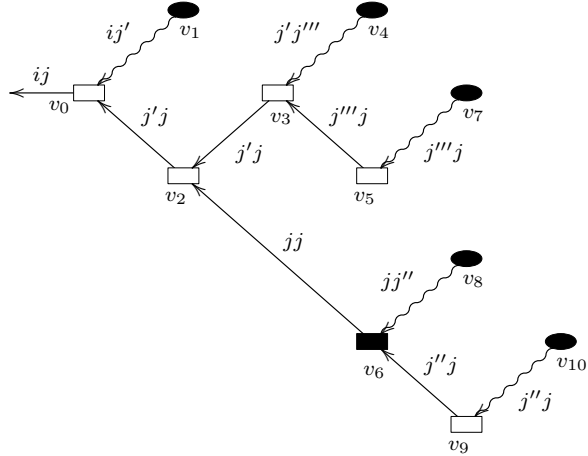

If $s_{v}$ denotes the number of lines entering $v$ and $s_v^\eta$ the number of lines of type $\eta$ entering 
$v$, we have the constraints $s_{v}\in\{0,1,2\}$ and $s_v^\eta\in\{0,1\}$. Moreover: $s_v=1\Rightarrow s_v^\mathfrak{M}=1$; 
$s_v=2\Rightarrow s_v^K=1$. If $s_v=1$ and $v_1$ is the node immediately preceding $v$ on has $i_v=i_{v_1}$ and $j_v=j_{v_1}$.
If $s_v=2$, let $v_1,v_2$ be the two nodes immediately preceding $v$; 
if $s_v^{\mathfrak M}=1$, with no loss of generality we assume that $v_1$ is of type $\mathfrak M$ (so that $v_2$ is of type $K$);
if $s_v^N=1$, with no loss of generality we assume that $v_1$ is of type $K$ (so that $v_2$ is of type $N$); in both cases
we impose the constraints that $i_{v}=i_{v_1}$, $j_{v}=j_{v_2}$ and $j_{v_1}=i_{v_2}$. 
Denoting by $v'$ the node immediately following $v$, we set $q_v:=s_{v'}^N$, and
$$p(v)= \sum_{w \in \mathcal{P}(v,v_0)} (m_{w}+q_w).$$
We are finally ready to define  the \emph{node factors} associated with the nodes:
\begin{equation} \label{eq:3.18}
A_{v}(\varphi) := \begin{cases}
1, & {\rm if}\quad \eta_{v}= N , \\
\displaystyle{-\a_v  \l_{i_v}^{-1}(\l_{i_v}/\l_{j_v})^{-m_v}(-1)^{s_v^N}}, & {\rm if}\quad \eta_{v}=K , \\
\displaystyle{\lambda_{j_{v}}\mathfrak{M}_{i_{v},j_{v}}^{(n_{v})}
(S^{p(v)}\varphi)} , &{\rm if}
\quad \eta_{v}=\mathfrak{M} .
\end{cases}
\end{equation}
Then, by iterating the graphical representation in Figures \ref{fig:Rappresentazione-grafica-di Gamma^n} 
and \ref{fig:Rappresentazione-grafica-di K^n}, we end up with trees like that in
Figure \ref{fig:Esempi-di-alberi 1 e 2} for $i\neq j$; note that
the end-nodes are all of type $\mathfrak{M}$. If $i=j$ the only difference is that the special node $v_0$ is of type $N$.

With the definitions above, we denote by $\Theta_{n,\eta,(i_v,j_v)}^{**}$ the set of labelled trees such that
$\sum_{v\in N(\theta)}n_v=n$, $\eta_{v_0}=\eta$, and the constraints and properties described above.
Then it is straightforward to prove by induction that 
\begin{equation} \label{eq:3.20}
\nu_{i}^{(n)}(\varphi)=\sum_{\theta\in\Theta^{**}_{n,N,(i,i)}} {\rm Val}(\theta) , \qquad
k_{i,j}^{(n)}(\varphi) =\sum_{\theta\in\Theta^{**}_{n,K,(i,j)}} {\rm Val} (\theta),
\end{equation}
where
\begin{equation} \label{eq:3.21}
{\rm Val}(\theta) := \prod_{v\in V(\theta)} A_{v} .
\end{equation}
Given $\bar\theta\in  \Theta^{**}_{n,\eta,(i,j)}$, we let $\mathcal{T}(\bar\theta)$ be the family of labelled trees differing from 
$\bar\theta$ just by the choice of the labels $\{m_v\}$. Then, 
using \eqref{e3.14}, it is easy to see that 
$$\sum_{\theta\in \mathcal{T}(\bar\theta)}|{\rm Val}(\theta)|\le \Big(\frac{\l_+}{1-\l_+^{-2}}\Big)^{N_i^K(\bar\theta)}
C_4^{E(\bar\theta)}\mu^{[(2n-E(\bar\theta))/3]},$$
which immediately implies that 
\begin{equation} \nonumber
\left| k_{i,j}^{(n)}(\varphi) \right|, \left| \nu_{i}^{(n)}(\varphi) \right| \le C_5^n\mu^{[(2n-1)/3]},
\end{equation}
for a suitable constant $C_5>0$. Therefore, the radius of convergence in $\e$ of the series for $K(\f)$ and $N_\e(\f)$ 
is proportional to $\m^{-2/3}$, which allows us to fix eventually $\e=\mu$.

The Lyapunov exponents $\Lambda_{i}$ are the time average of the quantities $\log \lambda_{i}(\varphi)$.
However, if $S^{2\pi}_{\e}|\Omega$ denotes the restriction of $S^{2\pi}_{\e}$ on the attractor $\Omega$,
the dynamical system $(\Omega,S^{2\pi}_{\e}|\Omega)$ is conjugated to
an Asonov system and hence it is ergodic: therefore time-averaged observables are $\varphi$-independent.
Furthermore, there exists a unique SRB measure $\mu_{\varepsilon}$ such that 
\begin{equation} \label{eq:3.22}
\Lambda_{\pm}=\int {\rm d}\mu_{\varepsilon}(\varphi)\log\left(\lambda_{\pm}+\nu_{\pm}(\varphi)\right),
\qquad \Lambda_{0}=\int {\rm d}\mu_{\varepsilon}(\varphi)\log\left(1+\nu_{3}(\varphi)\right) .
\end{equation}
The measure $\mu_{\e}$ can be computed by reasoning as in \cite[Chapter 10]{GBG}. Let $\mathfrak{P}_{0}$ be
a Markov partition for $S$ on $\mathbb T^2$ and set $\mathfrak{P}_{\e}=(H(\mathfrak{P}_0),W(\mathfrak{P}_0))$.
Call $X_{\e}$ the symbolic code induced by the Markov partition $\mathfrak{P}_{\e}$ and
denote by $\underline{\sigma}$ the symbolic representation of a point ${\bf x} \in\Omega$, i.e. ${\bf x}=X_{\e}(\underline{\sigma})$.
Then the expansion rate of $S^{2\pi}_{\e}|\Omega$ along the unstable manifold of $X_{\e}(\underline{\sigma})$ is
${\rm e}^{A_{+}(\underline{\sigma})}$, where
$$ A_{+}(\underline{\sigma}) = \log \left( \l_{+}+\nu_{+}( X_{0}(\underline{\sigma}) ) \right)
\frac{|{\bf w}_{+}(X_{0}(\tau\underline{\sigma}))|}{|{\bf w}_{+}(X_{0}(\underline{\sigma}))|} , $$
with $\tau$ denoting the shift map and 
${\bf w}_{+}(\f)=(\mathds{1}+K(\f)){\bf y}_{+}$.
If $\mu_{\e}$ denotes the Gibbs distribution for the energy function
$A_{+}(\underline{\sigma})$ (see \cite[Chapter 5]{GBG}),
then the SRB distribution $\mu_{\rm SRB}$ for the system $(\Omega,S^{2\pi}_{\e}|\Omega)$ is
$\mu_{\e}$ and can be computed accordingly (see \cite[Chapter 6]{GBG}).

Moreover, by construction, $A_{+}(\underline{\sigma})$ is analytic in $\e$ and H\"older-continuous in $\underline{\sigma}$.
Therefore, for any H\"older-continuous function $F:\mathbb T^2\to \mathbb R$, the expectation value
\begin{equation} \nonumber
\mu_{\e}(F) := \int {\rm d}\mu_{\e}(\varphi) \, F(\f)
\end{equation}
is H\"older-continuous in $\e$ for $0<\e\le \e_0$. If $F$ is analytic in $\e$, then there exists a positive constant $\e_{F}$,
depending on $F$, such that $\mu_{\e}(F)$ is analytic for $|\e| < \e_{F} \le \e_0$.
In particular the Lyapunov exponents \eqref{eq:3.22} are analytic in $\e$ and, from \eqref{eq:3.22}, one finds
\begin{equation} \nonumber
\Lambda_{0}=\int {\rm d}\mu_{\varepsilon}(\varphi)\log\left(1+\nu_{3}(\varphi)\right)=
\int {\rm d}\mu_{\varepsilon}(\varphi)\left(\varepsilon\Gamma+O\left(\varepsilon^{2}\right)\right)=
\varepsilon\Gamma+ O\left(\varepsilon^{2}\right) .
\end{equation}
This completes the proof of Theorem \ref{prop:2}.

%

\section{Perspectives: transition from fractal to smooth(er) synchronization}\label{sec:4}

In this section, we discuss informally some of the consequences of our main theorem, and 
formulate a conjecture about the transition from fractal to smooth(er) behavior, which is suggested by our result.
From Theorem \ref{prop:1}  we know that the surface $\O$ of the attractor 
is H$\overset{..}{\mbox{o}}$lder continuous, but we do not have any control on its possible differentiability. This means that our attractor may be fractal, and we actually expect this to be the case
for $\e$ positive and small enough. An analytic estimate of the fractal dimension of the 
attractor in terms of the Lyapunov exponents is provided by the {\it Lyapunov dimension} $D_L$, which is defined as follows. Consider an ergodic dynamical system admitting an SRB measure on its attractor, 
and let $\L_1\ge \L_2\ge \cdots\ge \L_n$
be its Lyapunov exponents, counted with their multiplicities. Then, 
\begin{equation}
D_{L}=k+\frac{\underset{i=1}{\overset{k}{\sum}}\Lambda_{k}}{\left|\Lambda_{k+1}\right|},\label{eq:Lyapdim}
\end{equation}
where $1\le k\le n$ is the largest integer such that $\underset{i=1}{\overset{k}{\sum}}\Lambda_{i}\geq0$.
The \textit{Kaplan-Yorke conjecture} \cite{Farmer-Ott-Yorke, Kaplan-Yorke} states that $D_{L}$ coincides with the 
Hausdorff dimension of the attractor (also known as the {\it information dimension}, see, e.g., \cite[Chapt.5.5.3]{Ga_fl}
for a precise definition). In this section, we take $D_L$ as a heuristic estimate of the 
fractal dimension of the attractor, without worrying about the possible validity of the conjecture
(which has been rigorously proven only some special cases, see e.g. \cite{{Eck-Ruelle}}).

Specializing the expression of $D_L$ to our context, we find that, for $\e$ sufficiently small,
\be D_L=2+\frac{\L_++\L_0}{|\L_-|}=3+\e\frac{\G}{\log\l_+}+R_2(\e),\label{eq:fract}\ee
where $R_2(\e)=O(\e^2)$ is the Taylor remainder of order 2 in $\e$, which 
is computable explicitly in terms of the convergent expansion 
derived in the previous sections. Note that $\G<0$, so that $D_L=D_L(\e)$ is smaller than 3 (as 
desired) and is decreasing in $\e$, for $\e$ small. Therefore, combined with the Kalpan-Yorke conjecture, (\ref{eq:fract}) suggests that the attractor is 
fractal for $\e$ small, and its fractal dimension decreases (as expected) by increasing the 
strength $\e$ of the dissipative interaction.

\begin{figure}[!h]
	\centering \setlength{\unitlength}{1.5mm}
	\begin{picture}(100,30)
	\put(0,0){\includegraphics[width=7cm]{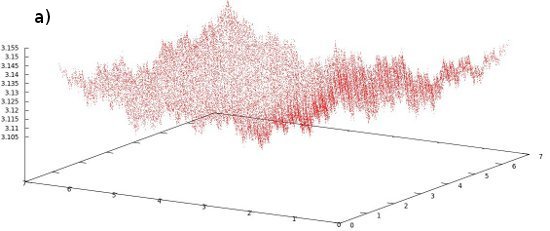}}
	\put(50,0){\includegraphics[width=7cm]{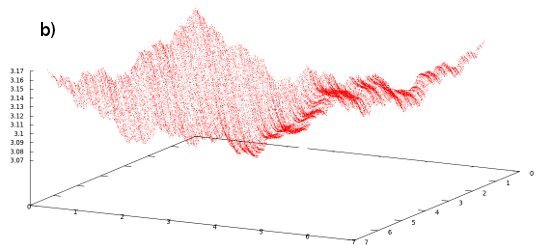}}
	\end{picture}
     \caption{\label{fig: f=0} $f=0$ and $g(\varphi,w,t)=\mbox{sin}(w-t)+\mbox{sin}(\varphi_{2}+w+t)$.
     a): $\varepsilon=0.05$. b): $\varepsilon=0.1$.}
	\label{fig1}
\end{figure}

\begin{figure}[!h]
	\centering \setlength{\unitlength}{1.5mm}
	\begin{picture}(100,30)
	\put(0,0){\includegraphics[width=7cm]{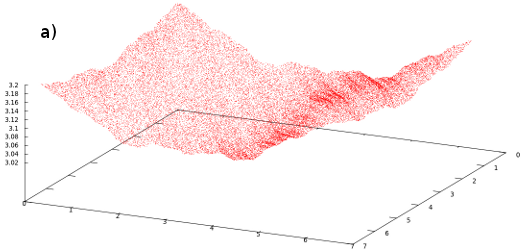}}
	\put(50,0){\includegraphics[width=7cm]{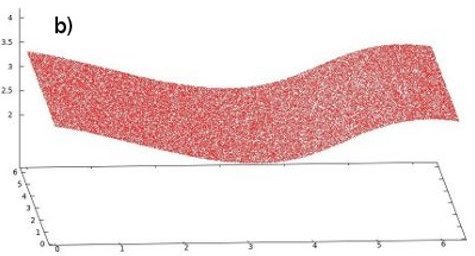}}
	\end{picture}
	\caption{\label{fig: f=0trans} $f=0$ and $g$ as in Fig. \ref{fig1}.
	a): $\e=0.153\simeq \e_c$.
	b): $\varepsilon=1$.}
	\label{fig2}
\end{figure}

It is now tempting to extrapolate \eqref{eq:fract} to larger values of $\e$ (possibly beyond the 
range of validity of Theorem \ref{prop:2}), up to the point where, possibly, the relative ordering 
of $\L_0$ and $\L_+$ changes. In the simple case that $f=0$ (which is the case considered in \cite{GGG}), the Lyapunov exponents $\L_\pm$ are independent of $\e$: $\L_+=-\L_-=\log\l_+$. Therefore, on the basis of \eqref{eq:fract}, we conjecture that by increasing $\e$ the Hausdorff 
dimension of the attractor decreases from $3$ to $2$ until $\e$ reaches the critical value 
$\e_c$, where $\L_0(\e)\big|_{\e=\e_c}=\L_-$. Formally, this critical point is $\e_c=(\log\l_+)/(-\G)+$(higher orders), the higher orders being computable via the expansion described in the previous sections.  For $\e\ge \e_c$, we expect the attractor to be a smooth manifold of dimension 
two. The transition is illustrated in Fig.\ref{fig1} and \ref{fig2} for the simple case that $f=0$ and 
$g(\varphi,w,t)=\mbox{sin}(w-t)+\mbox{sin}(\varphi_{2}+w+t)$, in which case the expected critical point is $\e_c\simeq 0.153$. 

\begin{figure}[!h]
	\centering \setlength{\unitlength}{1.5mm}
	\begin{picture}(100,30)
	\put(0,0){\includegraphics[width=7cm]{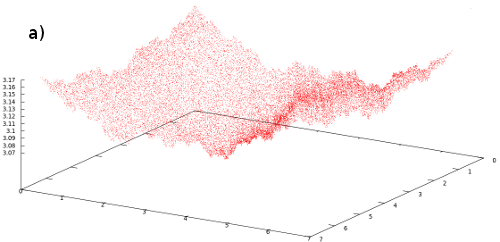}}
	\put(50,0){\includegraphics[width=7cm]{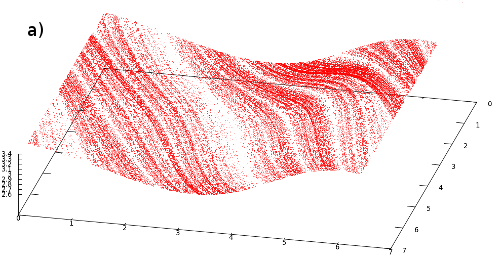}}
	\end{picture}
	\caption{$\boldsymbol{f}=(f_{1},f_{2,}g)=(\mbox{cos}(\varphi_{1}+w+t),0,\mbox{sin}(w-t)+\mbox{sin}(\varphi_{2}+w+t))$.
	a): $\varepsilon=0.1$. b): $\varepsilon=0.8$.}
        \label{fig3}
\end{figure}

If $f\neq0$, on the basis of numerical simulations, the attractor does not seem to display a transition from a fractal set to 
a smooth manifold. Still,  for suitable choices of $f\neq0$, we expect the attractor to display 
a ``first order phase transition'', located at the value of $\e$ where $\L_0(\e)=\L_-(\e)$, to be called 
again $\e_c$. At $\e=\e_c$, the 
derivative of the 
Hausdorff dimension of the attractor with respect to $\e$ is expected to have a jump. 
A possible scenario is that the attractor is fractal both for $\e<\e_c$ and for $\e>\e_c$, 
but it is ``smoother'' at larger values of $\e$, in the sense that its closure may be a regular, smooth, manifold of dimension two. An illustration of this ``smoothing" mechanism is in Fig.\ref{fig3}.

It would be interesting to investigate the nature of this transition in a more quantitative way, 
by comparing a numerical construction of the attractor with the theory proposed here, 
obtained by extrapolating the convergent expansion described in this paper to 
intermediate values of $\e$. Such a comparison goes beyond the purpose of this paper, 
and we postpone the discussion of this issue to future research. 

\appendix

\section{Estimate  of $\boldsymbol{N_{i}^{1}(\theta)}$} \label{app:a}

One has $N_{i}^{1}(\theta)=0$ for $n=1$ and $N_{i}^{1}(\theta)\leq 1$ for $n=2$. 
Given a tree of order $n\geq3$ one proceed by induction.
Assume that $N_{i}^{1}(\theta')\leq(2n'-1)/3$ for all the trees $\theta'$ of order $n'\leq n-1$
and consider a tree $\theta$ of order $n$. Let $v_0$ be the special node of $\theta$, and
call $\theta_{1},\ldots,\theta_{s}$ the subtrees entering $v_{0}$, with $s \ge 1$.
If $v_0$ is a node of type 1, then $N_{i}^{1}(\theta)=1+\sum_{k=1}^{s} N_{i}^{1}(\theta_{k}) \leq1 + \sum_{k=1}^{s}
(2n_{k}-1)/3 \leq 1+(2(n-1)-s)/3=(2n+1-s)/3$, so that the bound follows for $s\ge 2$. If $s=1$, then the node
preceding $v_0$ cannot be of type 1. Call $\theta_{1}',\dots,\theta_{s'}'$ the
subtrees entering $v_{1}$, with $s'\ge 1$. Then one has $N_{i}^{1}(\theta)=1+\sum_{k=1}^{s'}
N_{i}^{1}(\theta_{k}') \leq1+(2(n-2)-s')/3=(2n-1-s')/3 \le (2n-1)/3$ for all $s'\geq 1$.
Finally if $v_0$ is not a node of type 1, then it has not to be counted and the argument follows by
using the inductive bounds for the subtrees entering $v_0$.

Moreover the bound in Lemma \ref{lem:2.1} is optimal. Indeed there are
trees $\theta$ of order $n$ such that $N_{i}^{1}(\theta)=(2n-1)/3$. 
Define recursively the level $\ell(v)$ of a node $v$ by setting $\ell(v)=0$
if $v$ is an end-node and $\ell(v)=p$ if at least one line entering
$v$ exits a node $w$ with level $\ell(w)=p-1$. Then consider a tree in which
all nodes except the end-nodes are circles (thais is of type 1 or 2) and have two entering lines
except those with level $1$ which have only one entering line;
see figure \ref{fig:3.8} for an example with $n=11$ ( \raisebox{1mm}{\xymatrix{*+<3.0pt>[F**:black]+[F**:white]\txt{}}}
means we can have any kind of square node). For such trees one has
$n=2^{k}-1+2^{k-1}$, where $k=\ell(v_0)$, $2^{k}-1$ is the number of internal nodes
and $2^{k-1}$ is the number of end-nodes. Hence $N_{i}^{1}(\theta)=2^{k}-1=(2n-1)/3$.

\begin{flushleft}
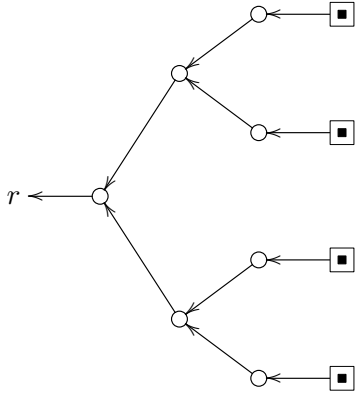
\begin{figure}
\noindent \raggedright{}\xymatrix@R=0.5cm{     &     &   &    *+[o][F]{}\ar@{<-}[r] &  *+<3.0pt>[F**:black]+[F**:white]\txt{}      \\
                                               &     &  *+[o][F]{}\ar@{<-}[ru]\ar@{<-}[rd]&   &  \\
                                               &     &   &    *+[o][F]{}\ar@{<-}[r] &  *+<3.0pt>[F**:black]+[F**:white]\txt{}      \\ 
                                               \hspace{3cm} {r}\ar@{<-}[r] & *+[o][F]{}\ar@{<-}[ruu]\ar@{<-}[rdd] &     &  &  \\ 
                                               &     &   &    *+[o][F]{}\ar@{<-}[r] &  *+<3.0pt>[F**:black]+[F**:white]\txt{}       \\ 
                                               &     &  *+[o][F]{}\ar@{<-}[ru]\ar@{<-}[rd]&   &  \\
                                               &     &   &    *+[o][F]{}\ar@{<-}[r] &  *+<3.0pt>[F**:black]+[F**:white]\txt{}    }
                                               \caption{A tree of order $n=11$ for which the bound on $N_{i}^{1}(\theta)$ is optimal. 
                                               \label{fig:3.8}}
\end{figure}

\par\end{flushleft}

{\bf Acknowledgments} G.G. and A.G. acknowledge financial support from the PRIN National Grant {\it Geometric and analytic theory of Hamiltonian systems in finite and infinite dimensions}. 


\end{document}